\documentclass[a4paper]{article}

\usepackage{INTERSPEECH2022}
\usepackage{amsmath,graphicx, booktabs,multirow, float}
\usepackage{xcolor}
\DeclareMathOperator*{\argmax}{argmax}
\usepackage{cite}

\title{Internal Language Model Estimation Through Explicit Context Vector Learning for Attention-based Encoder-decoder ASR
}
\name{Yufei Liu$^{1,2}$, Rao Ma$^1$, Haihua Xu$^1$, Yi He$^1$, Zejun Ma$^1$,Weibin Zhang$^2$}
\address{
	$^1$ ByteDance AI LAB \\
	$^2$South China University of Technology, Guangzhou, China
}

\begin{document}

\maketitle
\begin{abstract}

An end-to-end (E2E) ASR model implicitly learns a prior Internal Language Model (ILM) from the training transcripts. To fuse an external LM using Bayes posterior theory, the log likelihood produced by the ILM has to be accurately estimated and subtracted. In this paper we propose two novel approaches to estimate the ILM based on Listen-Attend-Spell (LAS) framework. 
The first method is to replace the context vector of the LAS decoder at every time step with a vector that is learned with training transcripts. 
Furthermore,
we propose another method that uses a lightweight feed-forward network to 
directly map query vector to context vector in a dynamic sense.
Since the context vectors  are  learned by minimizing the perplexities on training transcripts,
and their estimation is independent of encoder output,
hence the  ILMs are accurately learned for both methods.
Experiments show that the ILMs achieve the lowest perplexity, indicating the efficacy of the proposed methods. In addition, they also significantly outperform the shallow fusion method, as well as two previously proposed ILM Estimation (ILME) approaches on several datasets.

\end{abstract}

\section{Introduction}\label{sec:intro}
			
End-to-end (E2E) Automatic Speech Recognition (ASR) models\cite{chiu2018state,graves2014towards,rao2017exploring,amodei2016deep} are becoming more and more popular due to 1) its success in achieving state-of-the-art results; and 2) its compactness in that both the acoustic and language models are jointly learned with a single network. One of the most popular E2E models is the Listen-Attend-Spell (LAS) model \cite{chan2015listen}, which is also called the attention-based encoder-decoder (AED) model. 
	
Though compact and effective in modeling, E2E models are inherently limited in taking full advantage of an external language model (LM) that is trained on a much larger text-only data. This can be explained by the Bayes probabilistic theory that governs speech recognition models. Given an acoustic feature sequence $X$ and the corresponding word/sub-word sequence $W$,  an E2E model directly learns the posterior $P(W|X)$. $P(W|X)$ can be decomposed into an acoustic model $P(X|W)$ and a language model $P_{\texttt{prior}}(W)$.  The language model $P_{\texttt{prior}}(W)$ is trained by using only the training scripts and thus is suboptimal. When more external text data is available, one can train a much more robust external language model $P_{\texttt{ext}}(W)$. To fuse $P(W|X)$ with the external LM $P_{\texttt{ext}}(W)$, the effect of the internal LM $P_{\texttt{prior}}(W)$ has to be removed first. 
	
However, the difficulty lies in that we cannot easily estimate $P_{\texttt{prior}}(W)$ in an E2E model. Many simplified methods just ignore the effect of $P_{\texttt{prior}}(W)$ , such as component fusion \cite{shan2019component}, cold fusion \cite{sriram2017cold},and shallow fusion \cite{stahlberg2018simple}. As the internal language model $P_{\texttt{prior}}(W)$ is implicitly retained in $P(W|X)$, the fusion process could be biased, yielding sub-optimal results in both intra- and inter-domain ASR scenarios.

Recently, Hybrid Autoregressive Transducer (HAT)~\cite{variani2020hybrid} and Density Ratio approaches~\cite{mcdermott2019density, zeyer2021librispeech, sugiyama2012density} were proposed as an extension of shallow fusion~\cite{mikolov2010recurrent,chorowski2016towards,hori2017multi,kannan2018analysis,toshniwal2018comparison}. The Density Ratio method uses a separate model trained with  training transcripts to approximate $P_{\texttt{prior}}(W)$. By contrast, HAT estimates the ILM by removing the effect of the encoder from the RNN-T network. Both methods have been shown to outperform shallow fusion, especially in cross-domain tasks. Inspired by HAT, Meng \textit{et al}. proposed an Internal Language Model Estimation (ILME) method~\cite{meng2021ilme} to estimate the prior for both the RNN-T and the AED models. 
	
Though yielding performance gains in~\cite{meng2021ilme},
zeroing out the encoder’s output may potentially lead to mismatch during inference.
To deal with the mismatch issue, Meng \textit{et al}. have proposed an ILM training method in~\cite{meng2021ilmt} to  update the model parameters engaged in predicting the ILM score more recently. Moreover, to relax the constraint of zeroing out the encoder's output, Zeineldeen \textit{et al}. proposed a series of improved ILM estimate methods~\cite{zeineldeen2021investigating}.


In this paper\footnote{This work was done when Yufei Liu was an intern in ByteDance. }, we propose two novel ILME methods to estimate the ILM under AED framework. One of the simpler methods is to explicitly treat the decoder as an LM that is responsible for the ILM score estimation. Different from previous works, we propose to learn such an ILM through optimizing the context vector with training transcripts after the normal ASR training. The advantage of the proposed method is the simplicity to interpret  and yet can be applied to 
different AED framework.
Alternatively, 
as the learned context vector is fixed during inference,
we propose another method that allows for different context vector at each decoding step. 
Specifically, we propose to use a lightweight feed-forward network to map 
the query vector to the context vector.


\section{Related work}~\label{sec:related-work}
Though the proposed work is inspired by diversified E2E ASR-based LM fusion works~\cite{shan2019component,sriram2017cold,gulcehre2015using}, as well as shallow fusion works~\cite{mikolov2010recurrent,chorowski2016towards,hori2017multi,kannan2018analysis,toshniwal2018comparison},
the most related ones are from \cite{meng2021ilme}, \cite{meng2021ilmt} and \cite{zeineldeen2021investigating} respectively.

Our first proposed ILME method is equivalent to the combination of works \cite{meng2021ilme}
and \cite{meng2021ilmt} for AED models. We freeze all the parameters of AED network, and employ the training transcripts to minimize the  perplexity of the decoder output, yielding a learned context vector from scratch. With the assistance of such a context vector, the decoder is more like a LM, namely an ILM. Here we have no zeroing  out operation, nor other unnecessary assumptions.

Recently, Zeineldeen \textit{et al}. proposed a series of improved ILME methods in~\cite{zeineldeen2021investigating}, of which, the most effective method employs  a ``Mini-LSTM" network to estimate decoding-synchronous context vector by taking decoding output as input. By a shallow contrast, our second ILME is very close to such a ``Mini-LSTM" method. However, the difference is decisive. To estimate context vector in this paper, we take query vectors in decoder, i.e, hidden state vectors, as input instead. Not only do the query vectors contain context information, and hence more ``internal", they also let us sufficiently take advantage of what has been learned by the AED decoder, and therefore only a lightweight feed-forward network is required to learn the mapping. By comparison, since the ``Mini-LSTM" takes decoder output as input as mentioned, it is decoupled with the existing decoding network, and more related to the prior density ratio method~\cite{mcdermott2019density}.

\section{Attention-based encoder-decoder ASR}~\label{sec:aed-intro}
		
	
The objective of  AED-based E2E model is to predict the posterior $P(\bf{W}|\bf{X};\theta^{\texttt{AED}})$ of a word sequence $\bf{W}$, given the input feature sequence $\bf{X}$. In an AED model \cite{chan2015listen}, the encoder learns to map the feature representation $\bf{X}$ to a higher level  representation $\bf{H}^{\texttt{enc}} \triangleq \{\bf{h}_1^{\texttt{enc}}...\bf{h}_t^{\texttt{enc}}... \bf{h}_T^{\texttt{enc}}\}$,  where the dimension and sequence length between $\bf{X}$ and $\bf{H}^{\texttt{enc}}$ are normally different due to the down-sampling operation. The attention network determines which subset of the sequence $\bf{H}^{\texttt{enc}}$ are to be attended, given the decoder’s hidden state representation $\bf{H}^{\texttt{dec}} \triangleq \{\bf{h}_1^{\texttt{dec}}...\bf{h}_i^{\texttt{dec}}... \bf{h}_I^{\texttt{dec}}\}$, where $\bf{h}_i^{\texttt{
dec}}$ acts as a query vector. That is
	\begin{align}
	\bf{a}_i &= \text{AttentionNet}(\bf{H}^{\texttt{enc}}, \bf{h}_i^{\texttt{dec}})~\label{eqn:atten-01} \\
	\bf{c}_i &= \sum_{t=1}^{T}a_{i,t}\bf{h}_t^{\texttt{enc}}~\label{eqn:atten-02}  
	\end{align}
	where $\bf{a}_i$ is the attention weighting vector at each step $i$ and $\bf{c}_i$ is the context vector that will be employed by the decoder to predict the next token. With the obtained context vector, the decoder proceeds as follows:
	\begin{align}
	&\bf{h}_i^{\texttt{dec}} = \text{DecoderRNN}(\bf{h}_{i-1}^{\texttt{dec}}, \text{Concat}(\bf{e}_{i-1}^{\texttt{dec}},\bf{c}_{i-1}))~\label{eqn:decoder} \\
	&\bf{z}_i = W_z\bf{h}_i^{\texttt{dec}} + b_z~\label{eqn:linear} \\
	&P(w_i|\bf{X},W_{<i}; \theta^{\texttt{AED}})= \text{Softmax}(\bf{z}_i)\label{eqn:token-post}
	\end{align}
	where $\bf{e}_{i-1}^{\texttt{dec}}$ is the embedding vector corresponding to the output token $w_{i-1}$. $w_{i-1}$ is normally a word-piece in practice. As a result, the word/token sequence posterior $P(\bf{W}|\bf{X};\theta^{\texttt{AED}})$ is obtained as the product of equation (\ref{eqn:token-post}).
	
	\section{Proposed method}\label{sec:proposed-method}
	\subsection{One-time context learning (OTCL) for ILME}~\label{sub:nonadaptive}
From what is formularized in Section~\ref{sec:aed-intro}, if each context vector $\bf{c}_i$ in Eq.~(\ref{eqn:atten-02}) is not estimated from the encoder’s output, the decoder itself can be seen as an LM. This motivates us to learn a context vector by only using  training transcripts. Specifically, during inference Eq.~(\ref{eqn:atten-02}) is rewritten as 
	\begin{align}
	\bf{c}_i &=\bf{c} ~\label{eqn:trans-context}, \forall{~i} 
	\end{align}
where $\bf{c}$ is a learned vector, and once it is learned it is kept fixed during inference. For brevity, we name it as one-time context learning (OTCL) based ILME. 
Notably, during  training of $\bf{c}$, all the other parameters of the AED model are kept fixed. The learned context vector, together with the previously learned decoder, forms the estimated ILM. 

	\subsection{Label-synchronous context learning (LSCL) for ILME}~\label{sub:adaptive}
In Section~\ref{sub:nonadaptive} a static context vector that is kept fixed during inference at each time step is learned. This may not be optimal. From Eq.~(\ref{eqn:decoder}), we can see that the context vector varies along with the decoding state $\bf{h}_i^{\texttt{dec}}$ at every decoding step. The actual internal LM may benefit from this variation. Thus, we propose another ILME method, namely label-synchronous context learning (LSCL) method, where $\bf{c}_{i-1}$ is allowed to change at each decoding step as it is shown in Eq.~(\ref{eqn:decoder}). However, as a pure LM, $\bf{c}_{i-1}$  is not allowed to depend on the output of the encoder.
 
To achieve the objective as mentioned, we propose to generate the context vector by using only the current decoder state $\bf{h}_i^{\texttt{dec}}$. Specifically, 
we use a nonlinear function $\bf{f}$ to do the mapping i.e.   $\bf{c_i}=f(\bf{h}_i^{\texttt{dec}})$.
As a result, our objective is to learn such a nonlinear mapping function, the learned mapping function, together with the decoder, forms our ILM. 
To make learning simpler, 
we propose to use a lightweight Feed-Forward Neural Network (FFNN) for $\bf{f}(\cdot)$, i.e.
	\begin{align}
	\bf{c}_i = \text{FFNN}(\bf{h}_i^{\texttt{dec}})~\label{eqn:adaptive}
	\end{align}

Training of the FFNN is similar to what is described in Section~\ref{sub:nonadaptive}. Once the training of the entire AED model is done. We continue to train the FFNN to minimize the perplexity on the training transcript. The parameters for the trained AED model are kept fixed during updating of the FFNN. 
 
As mentioned in Section~\ref{sec:aed-intro}, Zeineldeen \textit{et al}. recently proposed a similar context learning method, called ``Mini-LSTM" in \cite{zeineldeen2021investigating}. 
 While the commonality is that both methods have introduced an extra network
to estimate context vectors for ILME, namely, ``Mini-LSTM" versus ``FFNN",  the difference between 
two approaches are remarkable.
In \cite{zeineldeen2021investigating}, decoder's output ${ w_i\in \bf{W}}$ is used as input, while we use decoder's state vector $\bf{h}_{i}^{\texttt{dec}}$ instead. Since  $\bf{h}_{i}^{\texttt{dec}}$ has already embedded historical information, it is sufficient for us to model the mapping by a lightweight FFNN, reducing the risk of overfitting. Moreover, as we use hidden state vector, the FFNN is more coupled with the existing decoder, and hence it is more ``internal".
In contrast, the ``Mini-LSTM" in~\cite{zeineldeen2021investigating} is actually decoupled with the decoder, and it is more close to a density ratio method. More importantly, our experiments in what follows show the proposed method can
yield better results than the Mini-LSTM-based ILME method.

	\subsection{ILME-based Language model fusion}~\label{sub:fusion}
For ILME-based LM fusion, we need three scores to proceed during inference. The three scores are, the output from the AED ASR decoder $P(\bf{W}|\bf{X}; \theta^{\texttt{AED}})$, the score produced by the estimated ILM $P(\bf{W}; \theta^{\texttt{AED}})$, and finally the score calculated by an external LM $P(\bf{W}; \theta_{\texttt{ext}}^{\texttt{LM}})$. As a result, the final inference results are obtained with following equation:
	\begin{equation}
	\begin{aligned}
	\widehat{\bf{W}}   = \argmax_{\bf{W}}[\log P(\bf{W}|\bf{X};\theta^{\texttt{AED}}) &- \lambda_{\texttt{ILM}}\log P(\bf{W}; \theta^{\texttt{AED}})\\
	&+ \lambda_{\texttt{LM}} \log P(\bf{W}; \theta_{\texttt{LM}}^{\texttt{ext}})  ]
	\end{aligned}
	\label{eqn:lm-fusion}
	\end{equation}
	where $\lambda_{\texttt{ILM}}$ and $\lambda_{\texttt{LM}}$ are the estimated ILM and external LM weighting factors respectively.

	\section{Experiments}~\label{sec:exp}
	\vspace{-8mm}
	\subsection{Data}~\label{sub:data}
	
  To verify the efficacy of the proposed methods for both intra- and inter-domain LM fusion,
  we conduct experiments on 5 training data sets over three languages English, Japanese, and Mandarin respectively. Table~\ref{tab:data-desp} reports the details of training data distributions.
  There are three test sets. For English, we use Librispeech\cite{panayotov2015librispeech} \texttt{test-other} for both cross- and intra-domain tests. For Japanese, we use our in-house test data that contains 1538 utterances for intra-domain test. While for Mandarin, we have an in-house medical domain test set with 1092 utterances for cross-domain test. 
  

			\begin{table}[H]
		\centering
			\caption{Training data description: transcribed data for AED-based E2E ASR modeling and text data building LM for both intra- and cross-domain-based LM fusions.``Giga" refers to Gigaspeech, ``Libri" refers to Librispeech, while ``house" is for ``in-house".}~\label{tab:data-desp}
		\begin{tabular}{llllll}
			\toprule
			\multirow{2}{*}{Language} &\multicolumn{2}{c}{E2E-ASR data} & \multicolumn{2}{c}{LM data}& \multirow{2}{*}{Domain}\\ 
			& Source &  Hours  & Source & Words & \\ \midrule
		    \multirow{3}{*}{English} &house & 18k &\multirow{3}{*}{Libri.} &\multirow{3}{*}{853M} & cross \\
		                             &Giga.  & 10k  &  & &cross  \\
		                             &Libri.  & 960   &  & & intra   \\
		    \midrule
		    Japanese  & \multirow{2}{*}{house} & 100 &\multirow{2}{*}{house} &913M& intra \\
		    
		    Mandarin &  &100k & &4183M  & cross\\
		    \bottomrule
		\end{tabular}
	\end{table}	
	
	
	\subsection{Models and experimental setups}~\label{sub:model}
Three different LAS models with different encoders, i.e., BLSTM, Transformer~\cite{vaswani2017attention,li2019improving}, and Conformer\cite{gulati2020conformer} are trained for comparison. The BLSTM encoder is configured the same as in ~\cite{chan2015listen}. Together with the decoder, they form a conventional LAS model ~\cite{chan2015listen}. The Transformer's main parameters \{layer, dim, head\} are \{18, 512, 8\}, and the intermediate GLU layer size is 2048 with 0.1 dropout. The Conformer's parameters \{layer, dim, head\} are \{12, 512, 8\} and the intermediate SWISH layer size is also 2048 with 0.1 dropout. The convolution kernel size for the Conformer is 32. As for the decoder, we use the same architecture for different LAS models. The decoder is a 4 layers LSTM with 1024 hidden units. For LAS models trained on the English dataset, we use the Byte Pair Encoding (BPE) subword units with a vocabulary of 7000.
For Japanese, the BPE is 8516. For Chinese, 8046 Chinese characters are used as the modeling units. The model used for external LMs is a 3-layer LSTM with 4096 units per layer. The feed-forward neural network in LSCL-ILME mapping is a fully connected 3 layers network with 512 units per layer. RELU is applied to the first two layers as the activation function. Both the learnable vector and the feed-forward network are trained with the initial learning rate of 0.001 and decay to the final learning rate of 0.0001 over 10000 steps. All the LM fusion parameters $\lambda_{\text{LM}}$ and $\lambda_{\text{ILM}}$ were tuned using grid search. Overall experiments are conducted on the  Lingvo~\cite{shen2019lingvo} platform.
	
\subsection{Results}~\label{sub:results}
\vspace{-8mm}
	\subsubsection{Cross-domain LM fusion}~\label{subsub:cross-domain}
Table~\ref{tab:cross-on-libri} reports the Word Error Rates (WERs) of the proposed methods for cross-domain LM fusion. The Librispeech \texttt{test-other} test data was used as the target domain, while the source ASR models were trained using the 18k-hours in-house English data.
	\begin{table}

		\centering
		\caption{WER(\%) on Librispeech \texttt{test-other}. Three AED models were trained with the 18k-hours in-house English data. Different fusion methods, including no fusion (None), Shallow Fusion (SF), Zero-out context~(Zero), Mini-LSTM~(LSTM), the OTCL-ILME (OTCL), and the LSCL-ILME (LSCL) are compared.}
		\label{tab:cross-on-libri}
		\resizebox{0.98\columnwidth}{!}{
			\begin{tabular}{lllllllll}
				\toprule
				\multicolumn{2}{c}{Encoder type}   &  None & SF &Zero &LSTM &OTCL &LSCL \\
				\midrule
				\multirow{3}{*}{BLSTM} &WER& 10.35&8.75&7.97&7.68&7.56&\textbf{6.88} \\
				&$\lambda_{\text{LM}}$&0.0&0.15&0.25&0.3&0.35&0.35 \\
				&$\lambda_{\text{ILM}}$&0.0&0.0&0.1&0.15&0.15&0.25 \\\midrule
				\multirow{3}{*}{Transformer}& WER&8.94&7.44&7.11&6.35&6.29&\textbf{5.99} \\
				&$\lambda_{\text{LM}}$&0.0&0.15&0.25&0.3&0.4&0.4\\
				&$\lambda_{\text{ILM}}$&0.0&0.0&0.05&0.15&0.2&0.25\\\midrule
				\multirow{3}{*}{Conformer}&WER&8.96&7.61&7.61&6.85&6.98&\textbf{6.41}\\
				&$\lambda_{\text{LM}}$&0.0&0.1&0.1&0.15&0.25&0.25 \\
				&$\lambda_{\text{ILM}}$&0.0&0.0&0.0&0.15&0.1&0.15\\
				\bottomrule
			\end{tabular}
		}
	\end{table}
As can be seen from Table~\ref{tab:cross-on-libri}, the proposed methods, particularly the LSCL-ILME method, achieve consistently the lowest WER with different encoders. Though the proposed OTCL-ILME method is very simple, it achieves comparable results with the Mini-LSTM method. The Zero-out method works slightly better than the shallow fusion method. Finally, compared with the shallow fusion method, the proposed LSCL-ILME can achieve up to 22\% relative WER reduction.


	Table~\ref{tab:cross-on-medical} reports the Character Error Rates~(CERs) on our in-house medical data set. The LAS model was trained on the 100k-hours Mandarin data. Due to space limitation, we only report the fusion results with the LAS model using Transformer encoder. From Table~\ref{tab:cross-on-medical}, the proposed LSCL-ILME fusion method achieves the best performance, making 28.57\% relative CER reduction over what is obtained without LM fusion, while making 19.05\% relative CER reduction over the shallow fusion result. More interestingly, its CER is 12.72\% better than the Mini-LSTM's. We can notice that Zero-out method~\cite{meng2021ilme} achieves no CER improvement over shallow fusion method under the Transformer-encoder-based LAS ASR framework. More details about this will be described in Section~\ref{sub:ablation}.
		\begin{table}[H]
		\centering
		\caption{CERs(\%) on the in-house Chinese medical test data.}
		\label{tab:cross-on-medical}
		\resizebox{0.98\columnwidth}{!}{
			\begin{tabular}{lllllllll}
				\toprule
				\multicolumn{2}{c}{Encoder type}   &  None & SF &Zero &LSTM &OTCL &LSCL \\
				\midrule
				\multirow{3}{*}{Transformer} &WER&6.72&5.93&5.93&5.50&5.50&\textbf{4.80} \\
				&$\lambda_{\text{LM}}$&0.0&0.15&0.15&0.35&0.35&0.45 \\
				&$\lambda_{\text{ILM}}$&0.0&0.0&0.0&0.25&0.25&0.4 \\
				\bottomrule
			\end{tabular}
		}
	\end{table}


\subsubsection{Intra-domain LM fusion}~\label{subsub:intra-domain}

Table~\ref{tab:cross-on-jp} reports CERs of our Japanese ASR system trained with 100 hours of data. 
Like Table~\ref{tab:cross-on-medical}, we only report the results of the Transformer-encoder-based LAS model.
From Table~\ref{tab:cross-on-jp}, the proposed LSCL-ILME method again produces the best result of the overall fusion methods. The improvement is smaller compared with what is achieved in Table~\ref{tab:cross-on-medical}.
This is understandable since the training data is rather small, with only 100 hours,
as a result, the influence of ILM should be  minor. However,
we can notice though the training data is rather small ILME-based LM fusion methods are consistently effective, yielding improved results over the conventional shallow fusion  result in Table~\ref{tab:cross-on-jp}.

		\begin{table}[H]
		\centering
		\caption{CERs(\%) on the in-house Japanese test data.}
		\label{tab:cross-on-jp}
		\resizebox{0.98\columnwidth}{!}{
			\begin{tabular}{lllllllll}
				\toprule
				\multicolumn{2}{c}{Encoder type}   &  None & SF &Zero &LSTM &OTCL &LSCL \\
				\midrule
				\multirow{3}{*}{Transformer} &WER&24.64&23.83&23.66&22.84&23.37&\textbf{22.77} \\
				&$\lambda_{\text{LM}}$&0.0&0.10&0.10&0.10&0.10&0.15 \\
				&$\lambda_{\text{ILM}}$&0.0&0.0&0.05&0.20&0.05&0.15 \\
				\bottomrule
			\end{tabular}
		}
	\end{table}

Table~\ref{tab:libri-intra} reports the WERs of various intra-domain LM fusion methods. The source E2E models were trained on the 960-hours Librispeech data set and evaluated on the \texttt{test-other} test set. 
From Table~\ref{tab:libri-intra}, both the proposed methods win an obvious margin over the Zero-out and the shallow fusion methods.
Meanwhile, the proposed LSCL achieves better results over the Mini-LSTM method.

	\begin{table}[H]
		\centering
		\caption{WERs(\%) on the Librispeech \texttt{test-other} test set with different intra-domain LM fusion methods.}
		\label{tab:libri-intra}
		\resizebox{0.98\columnwidth}{!}{
			\begin{tabular}{lllllllll}
				\toprule
				\multicolumn{2}{c}{Encoder type}   &  None & SF &Zero &LSTM &OTCL &LSCL \\
				\midrule
				\multirow{3}{*}{BLSTM} &WER&7.13&6.22&5.44&5.17&5.18&\textbf{5.16} \\
				&$\lambda_{\text{LM}}$&0.0&0.15&0.35&0.55&0.55&0.55 \\
				&$\lambda_{\text{ILM}}$&0.0&0.0&0.2&0.4&0.4&0.4 \\\midrule
				\multirow{3}{*}{Transformer}& WER&7.6&7.06&6.98&\textbf{6.51}&6.71&6.63\\
				&$\lambda_{\text{LM}}$&0.0&0.1&0.25&0.25&0.2&0.2\\
				&$\lambda_{\text{ILM}}$&0.0&0.0&0.15&0.15&0.15&0.1\\\midrule
				\multirow{3}{*}{Conformer}&WER&6.3&5.8&5.48&5.19&5.19&\textbf{5.11}\\
				&$\lambda_{\text{LM}}$&0.0&0.1&0.2&0.3&0.35&0.35 \\
				&$\lambda_{\text{ILM}}$&0.0&0.0&0.05&0.25&0.3&0.3 \\
				\bottomrule
			\end{tabular}
		}
	\end{table}

\subsection{Ablation}~\label{sub:ablation}
\vspace{-8mm}
\subsubsection{ Perplexity  of internal LM}~\label{subsub:perplexity}
One way to evaluate the efficacy of different ILME methods is to calculate the perplexity of the estimated ILM on the training text. 
To this end, we train the ILMs on Librispeech training transcript data.
Specifically, we divide the entire transcript into 2 parts, 90\% of utterances for training and 10\% utterances for validation.
Table~\ref{tab:ilme-ppl} presents the perplexity of LAS models with different encoders and different ILME methods.

			\begin{table}[H]
		\centering
			\caption{Perplexity of different LAS models and different internal language modeling methods evaluated on the held-out training transcript. ILM is trained on Librispeech data. }~\label{tab:ilme-ppl}
		\begin{tabular}{llll}
			\toprule
			Encoder Type     & BLSTM  & Transformer & Conformer \\ \midrule
			Zero-out     & 387  &  6247           & 3271        \\ 
			Mini-LSTM       & 240  &   460          & 477        \\  
			OTCL-ILM & 266  &   528             & 563        \\ 
			LSCL-ILM  & \textbf{235}  &   \textbf{428}             & \textbf{463}        \\ \bottomrule
		\end{tabular}
	\end{table}
	
\begin{figure}[!htb]
		\minipage{0.17\textwidth}
		\includegraphics[width=\linewidth]{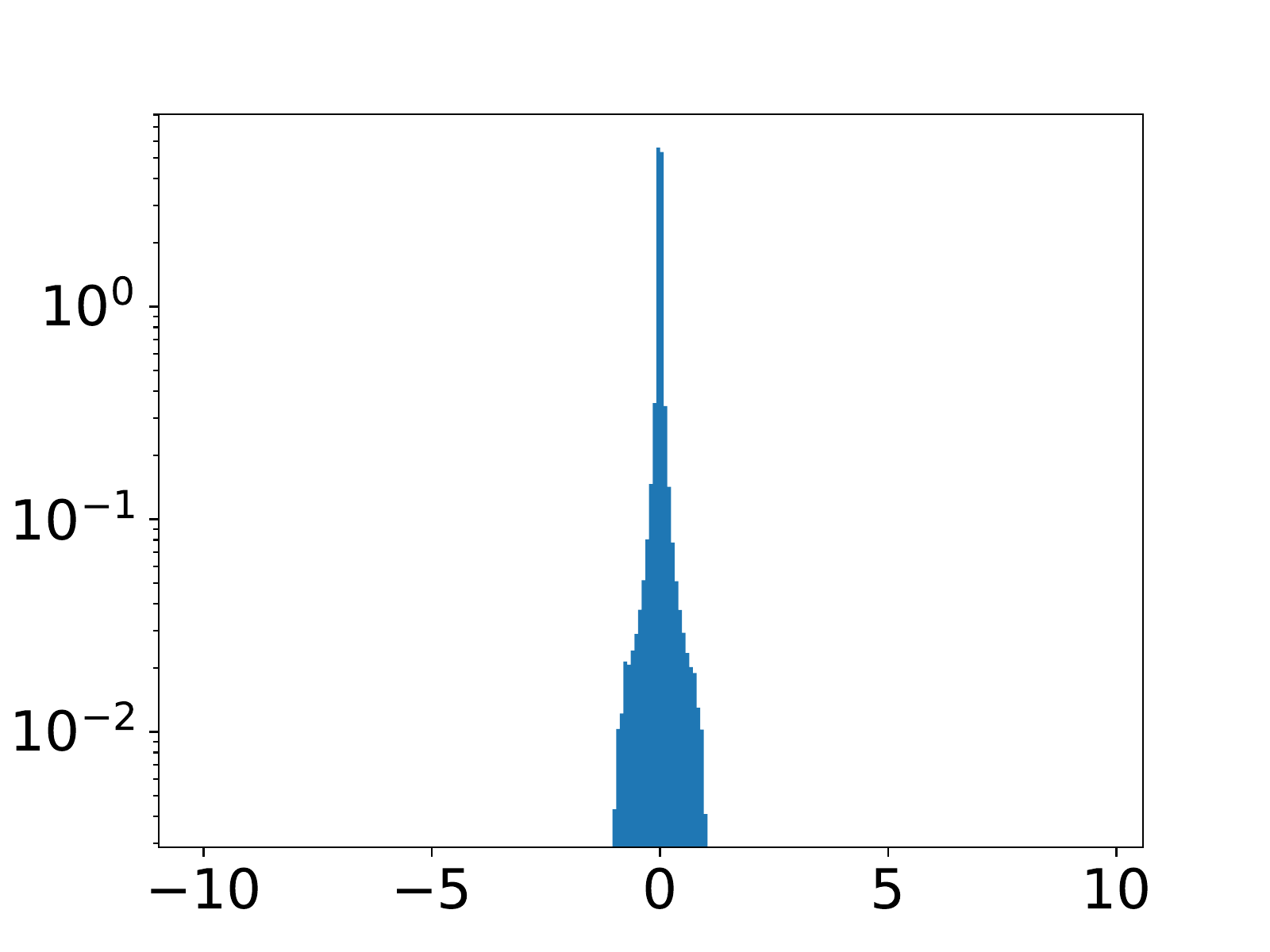}
		\centerline{(a) BLSTM}
		\endminipage
		\minipage{0.17\textwidth}
		\includegraphics[width=\linewidth]{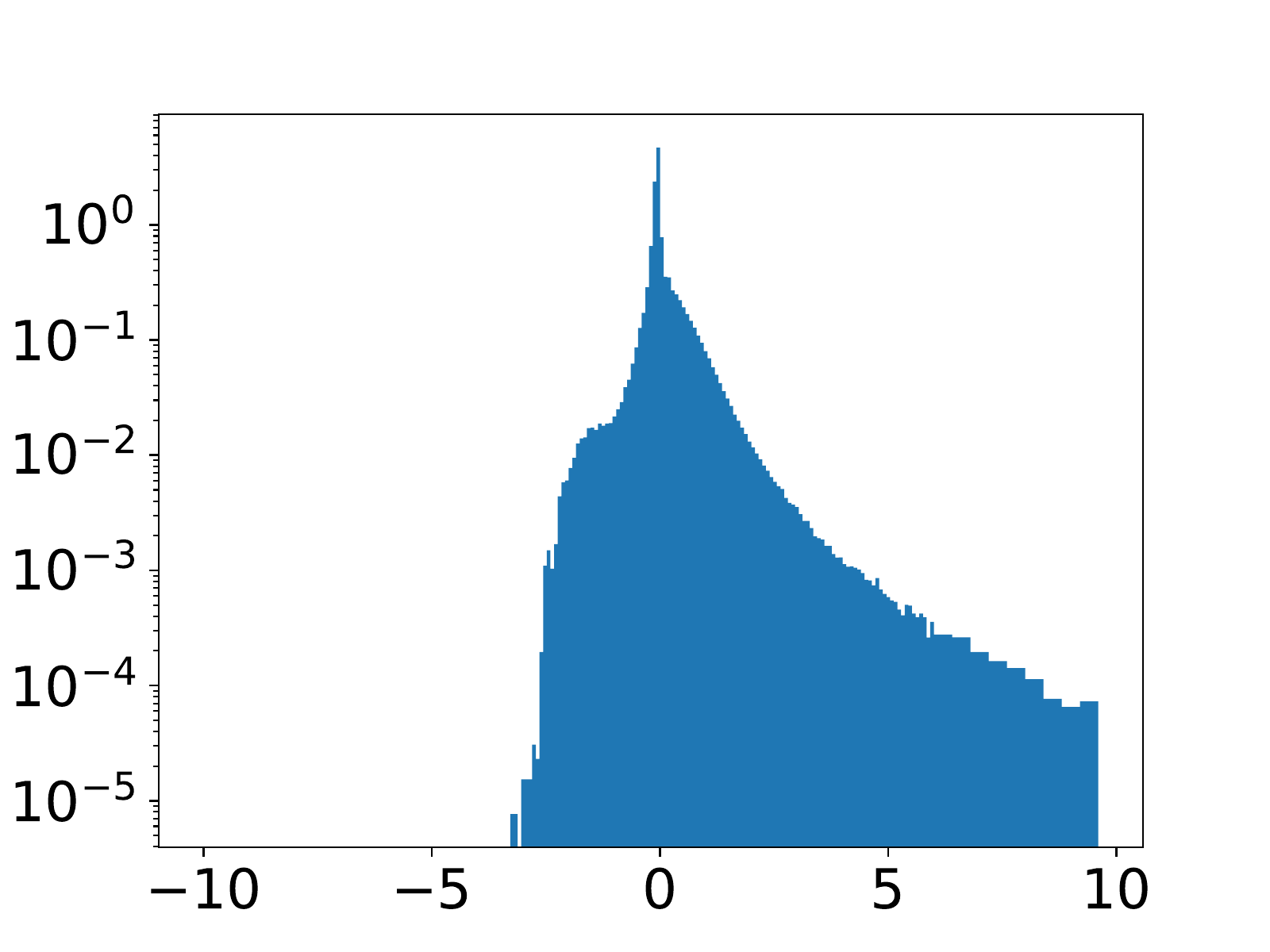}
		\centerline{(b) Conformer}
		\endminipage
		\minipage{0.17\textwidth}%
		\includegraphics[width=\linewidth]{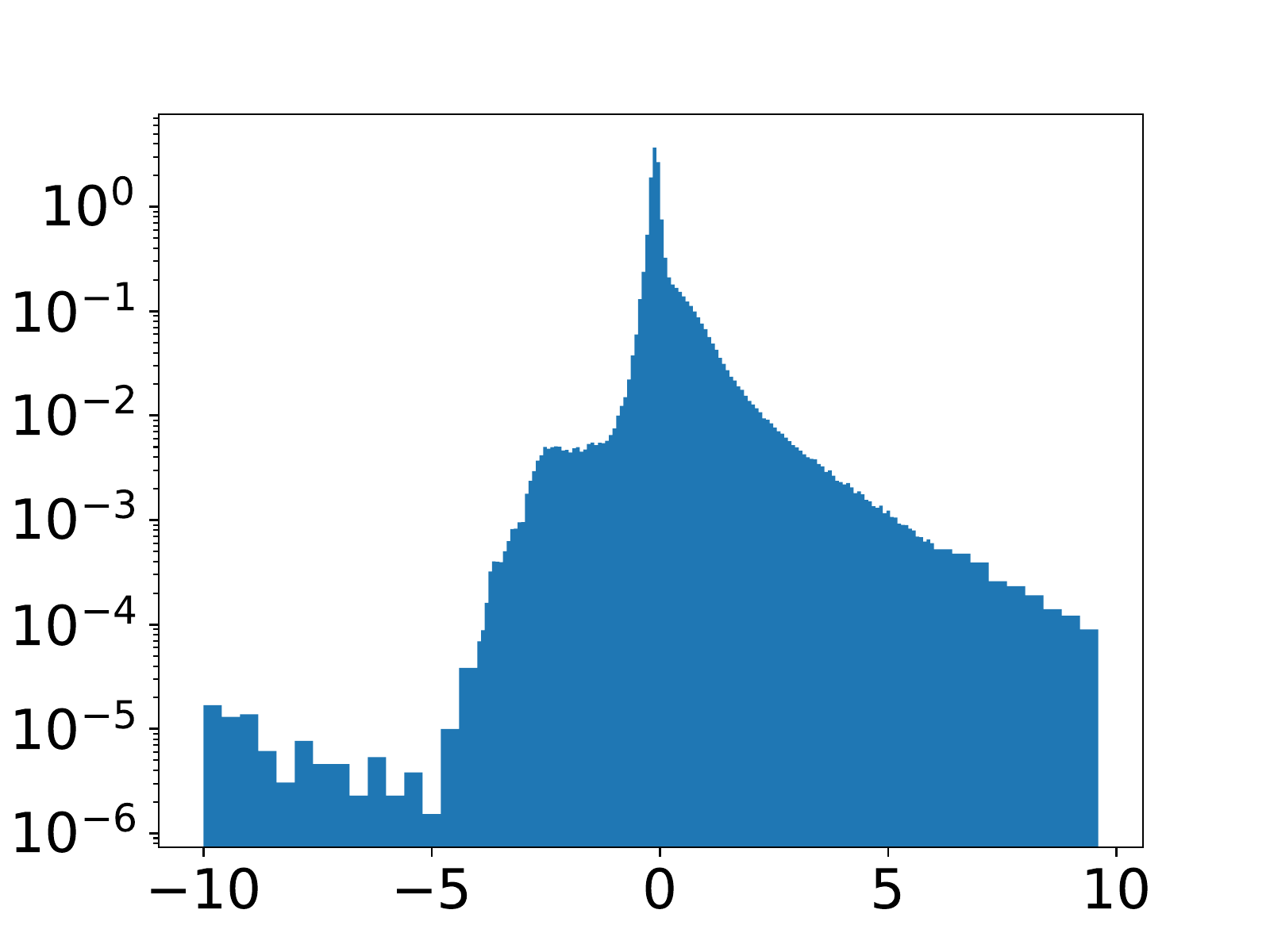}
		\centerline{(c) Transformer}
		\endminipage
		\caption{The numeric distribution of context vector from three different encoders.}
		\label{fig:dist-of-encoder-output}
	\end{figure}

From Table~\ref{tab:ilme-ppl}, we observe that the proposed LSCL-ILME method yields consistently the lowest perplexity among all methods. The performance is closely followed by the Mini-LSTM method. However, the perplexity of the Zero-out method~\cite{meng2021ilme} is very large, especially when it comes to Transformer or Conformer encoders. These abnormal perplexities explain why Zero-out method leads to rather poorer LM fusion results in the case of using Transformer and Conformer decoders. From Table~\ref{tab:cross-on-libri} to Table~\ref{tab:libri-intra}.
Directly zeroing-out context vector as an ILME method is probably not sensible. Therefore, we are curious about what context vector looks like for the three encoders. Figure~\ref{fig:dist-of-encoder-output} plots the distributions of context vectors from three different encoders.
																
 We can see that the distribution of context vector from the BLSTM’s encoder is symmetric. In addition. all values fall in the range of [-1,1]. Therefore, a zero-context vector is a reasonable assumption if the BLSTM is used as the encoder.  On the other hand, the distribution of context vector for both the Transformer and the Conformer encoders are rather ``messy". The dynamic range is also relatively large, leading to an ``unpredictable" context vector. Thus zeroing-out the context vector leads to very large perplexities due to assumption mismatch.
\subsubsection{Transformer Decoder}~\label{subsub:tsf-decoder}
So far we conduct all experiments using LAS ASR framework. A good ILME method should be free of specific decoder architecture. Table~\ref{tab:transformer-on-libri} reports our ILME performance on Librispeech \texttt{test-other} test set, where 
    the experiments are performed using Espnet\cite{watanabe2018espnet} instead\footnote{Source code: https://github.com/victor45664/espnet/tree/ilme}, and the model has a Conformer encoder and a Transformer decoder\cite{kim2017joint,zhou2018syllable}.
    The intra-domain ASR is trained with 960 hours of Librispeech data,  while the cross-domain ASR is trained with 10k hours of  Gigaspeech\cite{chen2021gigaspeech} data.
	\begin{table}[H]
		\centering
		\caption{WERs(\%) of the Librispeech \texttt{test-other}  with Transformer decoder using proposed ILME methods.}
		\label{tab:transformer-on-libri}
		\resizebox{0.98\columnwidth}{!}{
			\begin{tabular}{lllllllll}
				\toprule
				\multicolumn{2}{l}{Encoder type}   &  None & SF&OTCL &LSCL \\
				\midrule
				\multirow{3}{*}{Intra domain} &WER&6.0&4.7&4.37&\textbf{4.35} \\
				&$\lambda_{\text{LM}}$&0.0&0.7&0.9&0.9 \\
				&$\lambda_{\text{ILM}}$&0.0&0.0&0.6&0.6 \\\midrule
				\multirow{3}{*}{Cross Domain}& WER&7.9&5.56&4.89&\textbf{4.87}\\
				&$\lambda_{\text{LM}}$&0.0&0.5&0.7&0.9\\
				&$\lambda_{\text{ILM}}$&0.0&0.0&0.9&0.9\\
				\bottomrule
			\end{tabular}
		}
	\end{table}
From Table~\ref{tab:transformer-on-libri}, the proposed method has achieved significant performance improvement over the shallow fusion method in either case.

	\section{Conclusion}~\label{sec:con}
	
    In this paper, we proposed two novel ILME methods by learning a static context vector or a mapping between the query vector and the context vector. Experiments on multiple datasets demonstrate the effectiveness of the proposed methods. Compared with shallow fusion and other previously proposed ILME methods, the methods proposed in this paper significantly reduce the error rate of the system. In the future, we would like to extend these methods to the recurrent neural network transducer-based ASR framework.

	\vfill\pagebreak
\clearpage
\bibliographystyle{IEEEtran}

\bibliography{refs}

\end{document}